\documentclass[useAMS,usenatbib]{mn2e}
\usepackage{aas_macros}
\usepackage{amsmath,amssymb}
\usepackage{bm}
\usepackage{color}
\usepackage[dvipdfmx]{hyperref}
\usepackage{graphicx}
\usepackage{enumerate}

\newcommand{\nnmb}{\nonumber\\}

\newcommand{\mr}[1]{\mathrm{#1}}
\usepackage{amsmath}	% required for `\align' (yatex added)

\begin{document}
\author[K. Sugimura et al.
]{K. Sugimura,$^1$\thanks{E-mail: sugimura@astr.tohoku.ac.jp} C. M. Coppola,$^{2,3}$
K. Omukai,$^1$
D. Galli$^3$ 
and  F. Palla$^3$ \\
$^1$Astronomical Institute, Tohoku
University, Aoba, Sendai 980-8578, Japan \\ 
$^2$Dipartimento di Chimica, Universit\`a degli Studi di Bari, Via Orabona 4, I-70126 Bari, Italy\\
$^3$INAF-Osservatorio Astrofisico di Arcetri, Largo E. Fermi 5, I-50125 Firenze, Italy
} 
\title[Role of the $\mr{H}_2^+$ channel
]{Role of the $\mr{H}_2^+$ channel in the primordial star formation 
under strong radiation field and  
the critical intensity for the supermassive star formation
}
\maketitle
\topmargin-1cm

\begin{abstract}
We investigate the role of the $\mr{H}_2^+$
channel on H$_2$ molecule formation during the collapse of primordial gas clouds 
immersed in strong radiation fields which are assumed to have the shape of a 
diluted black-body spectra with temperature $T_\mr{rad}$. 
Since the photodissociation rate of $\mr{H_2^+}$ depends on its level population, 
we take full account of the vibrationally-resolved $\mr{H}_2^+$ kinetics.
We find that in clouds under soft but intense radiation fields with 
spectral temperature $T_\mr{rad}\lesssim 7000\,\mr{K}$, 
the $\mr{H_2^+}$ channel is the dominant H$_2$ formation process.  
On the other hand, for harder spectra with $T_\mr{rad}\gtrsim 7000\,\mr{K}$,
the $\mr{H^-}$ channel takes over $\mr{H_2^+}$ in the production of molecular hydrogen. 
We calculate the critical radiation intensity needed for supermassive star
formation by direct collapse and examine its dependence on the $\mr{H_2^+}$ level
population.
Under the assumption of local thermodynamic equilibrium (LTE) level population, the critical intensity is
underestimated by a factor of a few 
for soft spectra with $T_\mr{rad}\lesssim 7000\,\mr{K}$.
For harder spectra, the value of the critical intensity 
is not affected by the level population of $\mr{H_2^+}$.
This result justifies previous estimates of the critical
intensity assuming LTE populations since radiation sources
like young and/or metal-poor galaxies are predicted to have rather hard spectra. 
\end{abstract}


%%%%%%%%%%%%%%%%%%%%%%%%%%%%%%%%%%%%%%%%%%%%%%%%%%%%%
%%%%%%%%%%%%%%%%%%%%%%%%%%%%%%%%%%%%%%%%%%%%%%%%%%%%%
\section{Introduction}
\label{sec:intro}
%%%%%%%%%%%%%%%%%%%%%%%%%%%%%%%%%%%%%%%%%%%%%%%%%%%%%
%%%%%%%%%%%%%%%%%%%%%%%%%%%%%%%%%%%%%%%%%%%%%%%%%%%%%

%importance of H2 formation
The collapse of primordial gas clouds can lead to the formation of
ordinarily-massive stars
\citep[$10-100~M_{\odot}$, hereafter Pop III stars; see, e.g.,][]{Yoshida:2008aa, Hosokawa:2011aa}
or supermassive stars \citep[$M \gtrsim 10^5~M_{\odot}$, hereafter SMS; see, e.g., ][]{Bromm:2003aa},
depending on the strength of the radiation fields in the ambient medium. 
Pop III stars form by $\mr{H}_2$ cooling in the absence of a strong
Lyman-Werner (LW) radiation field ($h\nu=11.2-13.6\,\mr{eV}$). Conversely,
in a cloud irradiated by an extremely strong LW radiation field,
$\mr{H}_2$ is completely destroyed, and cooling is suppressed.
In the latter case, if the cloud is in a halo with virial temperature higher than $\sim 10^4\,\mr{K}$, 
it can still collapse via atomic cooling \citep{Omukai:2001aa}, without experiencing major episodes of 
fragmentation
\citep{Bromm:2003aa,Regan:2009aa,Regan:2009bb,Inayoshi:2014aa,Latif:2014ab,Regan:2014aa,Regan:2014bb,Choi:2015aa,Becerra:2015aa},
and an embryonic protostar can grow rapidly to a SMS by subsequent accretion \citep{Hosokawa:2012aa,Hosokawa:2013aa,Sakurai:2015aa,Shlosman:2015aa}. The latter eventually collapses due to the post-Newtonian instability, 
leaving a black hole with $\sim 10^5 M_\odot$ \citep[see, e.g.,][]{Shapiro:1983aa}. This process, called 
``direct'', or ``monolithic'' collapse, leads to the formation of
%SMSs and Jcr
SMS remnants that are promising candidates
for being the seeds of the supermassive ($\gtrsim
10^{9}M_\odot$) black holes (SMBHs) observed at redshift $z\gtrsim 6$
\citep[see, e.g.,][]{Fan:2001aa,Mortlock:2011aa,Venemans:2013aa,Wu:2015aa}).

The feasibility of this scenario can be tested by comparing
the observed number density of the high-$z$ SMBHs $n_\mr{SMBH}$ with the theoretical predictions.
By considering the probability distribution of 
the LW intensity $J_{21}$ around halos\footnote{
We use 
the specific intensity at the center of the LW bands, 
$J_{21}\equiv J_{\nu}(h\nu=12.4\,\mr{eV})/10^{-21}\,\mr{erg}\,\mr{s^{-1}}
\,\mr{cm^{-2}}\,\mr{Hz^{-1}}\,\mr{sr^{-1}}$ to quantify 
the intensity of external radiation.},
$n_\mr{SMBH}$ can be estimated from
the critical value of LW intensity $J_{21}^\mr{cr}$ needed for the formation of SMS
 \citep{Dijkstra:2008aa,Agarwal:2012aa,Dijkstra:2014aa,Inayoshi:2015ab}.
\cite{SOI14} (hereafter SOI14) found $J_{21}^\mr{cr}\sim 1000$ for realistic young-galaxy spectra. 
Combined with the $J_\mr{21}$ probability distribution by \cite{Dijkstra:2014aa}, 
this gives $n_\mr{SMBH}\sim 10^{-10}~\mr{Mpc^{-3}}$, in rough agreement with the observed values.
However, the predicted $n_\mr{SMBH}$ is
still uncertain due to our poor knowledge of $J_{21}^\mr{cr}$ 
from insufficient modelling of the physical/chemical processes in
primordial gas clouds \citep[see, e.g.,][]{Glover:2015aa,Glover:2015ab},
and of the nature of the high-redshift sources responsible for this radiation.

%H2 formation processes
In the low-density primordial gas ($\lesssim 10^{8}~{\rm cm^{-3}}$), 
$\mr{H}_2$, the main coolant in the low temperatures regime,
is formed either via the $\mr{H^-}$ channel or via the $\mr{H_2^+}$ channel.
The former begins with the formation of $\mr{H}^-$ by radiative
attachment,
    \begin{align}
      {\rm H} + e \rightarrow {\rm H^-} + {\rm \gamma}\,,
      \label{eq:7}
    \end{align}
followed by $\mr{H_2}$ formation by associative detachment
    \begin{align}
      {\rm H^-} + {\rm H} \rightarrow  {\rm H_2} +  \mr{e} \,.
      \label{eq:8}
    \end{align}
Similarly, the $\mr{H_2^+}$ channel 
starts with $\mr{H}_2^+$ formation by radiative association
    \begin{align}
      {\rm H} + {\rm H^+} \rightarrow {\rm H_2^+} + {\rm \gamma}\,,
      \label{eq:16}
    \end{align}
followed by the charge transfer reaction
    \begin{align}
      {\rm H_2^+} + {\rm H} \rightarrow  {\rm H_2} +   {\rm H^+}\,.
      \label{eq:18}
    \end{align}
In each channel, the second step proceeds much faster than the first one and its rate determines the amount
of H$_2$ formation.
Since the rate coefficient for reaction~\eqref{eq:7} 
\citep[$k_\mr{ra}^\mr{H^-}\sim 6\times10^{-15}\,\mr{cm^{3}\, s^{-1}}$ at 8000~K;][hereafter GP98]{Galli:1998aa}
is about one order of magnitude larger than that for reaction~\eqref{eq:16}
($k_\mr{ra}^\mr{H_2^+}\sim 1.5\times10^{-16}\,\mr{cm^{3}\, s^{-1}}$ at 8000~K; GP98),  
the $\mr{H}^-$ channel is usually much more efficient than  that involving $\mr{H}_2^+$. 
 However, under a strong radiation field,
both channels become irrelevant due to the destruction of the 
intermediaries 
by the $\mr{H^-}$ photodetachment 
    \begin{align}
      {\rm H^-} + {\rm \gamma} \rightarrow  {\rm H} + e \,,
     \label{eq:2}
    \end{align}
and $\mr{H}_2^+$ photodissociation
    \begin{align}
{\rm H_2^+} + {\rm \gamma} \rightarrow       {\rm H} + {\rm H^+}\,.
\label{eq:3}
    \end{align}
    
Since the binding energy of $\mr{H}^-$ ($0.76\,\mr{eV}$) is smaller 
than that of $\mr{H}_2^+$ ($2.7\,\mr{eV}$ for the ground state), 
$\mr{H}^-$ is photodestroyed more easily than $\mr{H}_2^+$ and 
reaction~\eqref{eq:18} can become the dominant source of H$_2$ in the primordial gas.

%importance of H2+ level population
Here, it should be noted that the $\mr{H_2^+}$ photodissociation rate 
depends sensitively on the internal level population,
because $\mr{H}_2^+$ in high vibrational levels is much more vulnerable to photodissociation.  
As an example, the binding energy of $\mr{H}_2^+$ in vibrational levels $v=6\mr{\ and\ } 18$ is
only $1.2\mr{\ and \ } 0.003\,\mr{eV}$, respectively.
Thus, the level population
of $\mr{H_2^+}$ is determined by a complex interplay of reactions, involving 
formation (reaction~\ref{eq:16}), collisional dissociation
(reaction~\ref{eq:18}), photodissociation (reaction~\ref{eq:3}), collisional
excitation/de-excitation and radiative de-excitation.  Therefore in order to accurately
implement the effects of the $\mr{H}_2^+$ channel, a detailed calculation of
its level population is needed.

%IGM case
In the post-recombination era, $\mr{H_2}$
formation is primarily controlled by
the $\mr{H}_2^+$ channel for redshift $z>$100, owing to the suppression of
$\mr{H^-}$ by the cosmic microwave background (CMB)
radiation (see, e.g., GP98). However, 
the efficiency of the $\mr{H}_2^+$ channel depends strongly on its level population: 
if $\mr{H}_2^+$ is in the ground state, 
the $\mr{H_2^+}$ channel is so efficient as to produce most of the 
$\mr{H_2}$ molecules at a level $f_{H_2}\sim 10^{-4}$.
On the other hand, under local thermodynamic equilibrium  (LTE), the
$\mr{H_2^+}$ channel makes only a minor contribution to
$\mr{H_2}$, whose final abundance remains  limited to $f_{H_2}\sim 10^{-6}$.
Recently, \cite{Hirata:2006aa}, \cite{Coppola:2011aa} (hereafter C11) and \cite{lincei} 
have studied the chemistry of the early Universe computing the population 
of the vibrational levels of $\mr{H_2}$ and $\mr{H_2^+}$ following a state-to-state reaction kinetics, 
and found that $\mr{H_2^+}$ forms by reaction~\eqref{eq:16}
preferentially in excited states. 
As a result, ${\rm H_2^+}$ channel does not dominate the formation of $\mr{H_2}$, and 
the final H$_2$ abundance is limited by $\mr{H}^-$ at a level $f_{H_2}\sim 10^{-6}$.

%aim of this work
Although the efficiency of the $\mr{H}_2^+$
channel depends strongly on the $\mr{H}_2^+$ level population, 
the LTE rate \citep{Stancil:1994aa,Mihajlov:2007aa} has been widely used 
in studying primordial gas clouds without a real justification.  
An exception is \cite{Glover:2015aa} 
who calculated $J_{21}^\mr{cr}$ for 
black-body-type spectra with temperatures 
$T_\mr{rad}=10^4\,\mr{K}$  and $10^5\,\mr{K}$
under two assumptions for the $\mr{H_2^+}$ level population:
({\em i}\/) all the $\mr{H_2^+}$ is in the vibrational ground
state,
and ({\em ii}\/) all levels are in LTE.
Although the difference in $J_{21}^\mr{cr}$ in the two cases
is not significant, it is not yet clear
in what circumstances and to what extent the $\mr{H_2^+}$ channel can affect 
the evolution of primordial gas clouds.

%in this paper
In this paper, we compute the thermal and chemical evolution of 
 primordial gas clouds under a strong external radiation field
with $J_{21}$ around $J_{21}^\mr{cr}$
by computing the $\mr{H_2^+}$ vibrational level population, 
and assess the effect of $\mr{H}_2$ formation via the $\mr{H}_2^+$
channel.  We also determine the
critical intensity for supermassive star formation 
and examine its dependence on the $\mr{H_2^+}$ level population
by comparing our non-LTE results to those obtained 
in the LTE or ground state approximations.

%orgnization
The paper is organized as follows. 
In Sec.~\ref{sec:method}, we describe our model for collapsing primordial gas clouds. 
In Sec.~\ref{sec:result}, we present the result of our calculation. 
The implications and conclusions are described Sec.~\ref{sec:discussion}.
%%%%%%%%%%%%%%%%%%%%%%%%%%%%%%%%%%%%%%%%%%%%%%%%%%%%%
%%%%%%%%%%%%%%%%%%%%%%%%%%%%%%%%%%%%%%%%%%%%%%%%%%%%%
\section{Model}
\label{sec:method}
%%%%%%%%%%%%%%%%%%%%%%%%%%%%%%%%%%%%%%%%%%%%%%%%%%%%%
%%%%%%%%%%%%%%%%%%%%%%%%%%%%%%%%%%%%%%%%%%%%%%%%%%%%%
%%%%%%%%%%%%%%%%%%%%%%%%%%%%%%%%%%%%%%%%%%%%%%%%%%%%%
%%%%%%%%%%%%%%%%%%%%%%%%%%%%%%%%%%%%%%%%%%%%%%%%%%%%%
\subsection{Basic Equations}
\label{sec:1-zone}
%%%%%%%%%%%%%%%%%%%%%%%%%%%%%%%%%%%%%%%%%%%%%%%%%%%%%
%%%%%%%%%%%%%%%%%%%%%%%%%%%%%%%%%%%%%%%%%%%%%%%%%%%%%
%overview of the model
To follow the gravitational collapse of primordial gas clouds, we use
the one-zone model described in SOI14 \citep[see also][]{Omukai:2001aa}), updated as follows: 
the vibrational level population of $\mr{H_2^+}$
is resolved following C11; a part of the
chemical network is updated following \cite{Glover:2015aa,Glover:2015ab}.  
In our model, we compute physical quantities in the homogeneous central part 
of the self-similar solution of collapsing clouds
\citep{Penston:1969aa,Larson:1969aa,Yahil:1983aa}.  The qualitative
validity of the one-zone model has been confirmed recently by
three-dimensional hydrodynamical simulations
\citep{Shang:2010aa,Latif:2014ab}. Thanks to the substantial simplification
of gas dynamics, we can focus on the thermo-chemical processes in detail.

%brief review
According to the one-zone model, the evolution
of the gas density $\rho$ is modeled as
\begin{align}
\frac{\mr{d}\rho}{\mr{d}t}=\frac{\rho}{t_\mr{ff}}\,,
\end{align}
where $t_\mr{ff}=\sqrt{3\pi/32G(\rho+\rho_\mr{DM})}$ is the free-fall time,
$G$ the gravitational constant
and $\rho_\mr{DM}$ the dark matter density,
which is assumed to evolve following the solution of the spherical top-hat collapse model until it reaches the virial density.
The radius of the core is approximately given by the Jeans length $\lambda_\mr{J}=\sqrt{\pi k T_\mr{gas}/G\rho \mu m_\mr{p}}$,
where $m_\mr{p}$ is the proton mass, $\mu$ the mean molecular weight
and $k$ the Boltzmann constant.
The evolution of the gas temperature $T_\mr{gas}$ is determined by the energy equation,
\begin{align}
\frac{\mr{d}e_\mr{kin}}{\mr{d}t}=\Gamma_\mr{c} - \Lambda_\mr{rad} -\frac{\mr{d}e_\mr{int}}{\mr{d}t}\,,
\label{eq:14}
\end{align}
where $e_\mr{kin}=3k\,T_\mr{gas}/2\mu\, m_\mr{p}$ is the kinetic energy
of the gas per unit mass, $\Gamma_\mr{c}=(\rho\, k\,T_\mr{gas}/\mu\,
m_\mr{p}) \, (\mr{d}/\mr{d}t)\,(1/\rho)$ the compressional heating rate,
$\Lambda_\mr{rad}$ the net cooling/heating rate due to radiative
processes and $e_\mr{int}$ the internal energy of the gas per unit mass,
including both chemical energy and internal molecular energy.   In the
density and temperature range of our interest, the dominant cooling
processes are $\mr{H}_2$ \citep{Glover:2008aa}, updated according to
\cite{Glover:2015aa}, and Ly-$\alpha$ 
emission \citep{Anninos:1997aa}. Other radiative reactions or the time variation of
$e_\mr{int}$ hardly affect the evolution of $T_\mr{gas}$.  

%external radiation
We model the external radiation field as a diluted black-body with cut-off energy at
$13.6\,\mr{eV}$, specifying its temperature $T_\mr{rad}$ and
intensity $J_{21}$ at 12.4 eV.  
We limit our analysis to 
black-body spectra with various $T_\mr{rad}$ 
although realistic spectra of galaxies are more complex and not
well represented by a black-body spectrum
\citep[SOI14,][]{Agarwal:2015ab}.  SOI14 claimed that the ratio of the $\mr{H^-}$
photodetachment rate to the $\mr{H_2}$ photodissociation rate,
$k_\mr{pd}^\mr{H^-}/k_\mr{pd}^\mr{H_2}$, is a key parameter 
characterizing the hardness of the spectra, and
that typical young and/or
metal-poor galaxies have hard spectra with
$k_\mr{pd}^\mr{H^-}/k_\mr{pd}^\mr{H_2}\lesssim 10^3$, corresponding
to black-body spectra with $T_\mr{rad}\gtrsim2\times10^4\,\mr{K}$.
Nonetheless, we explore a broad range of black-body temperatures,
$T_\mr{rad}> 5\times10^3\,\mr{K}$, to mimic the evolution in a variety of environments possibly present
in the early Universe.

%%%%%%%%%%%%%%%%%%%%%%%%%%%%%%%%%%%%%%%%%%%%%%%%%%%%%
%%%%%%%%%%%%%%%%%%%%%%%%%%%%%%%%%%%%%%%%%%%%%%%%%%%%%
\subsection{Chemistry}
\label{sec:chemistry}
%%%%%%%%%%%%%%%%%%%%%%%%%%%%%%%%%%%%%%%%%%%%%%%%%%%%%
%%%%%%%%%%%%%%%%%%%%%%%%%%%%%%%%%%%%%%%%%%%%%%%%%%%%%
\begin{figure}
    \centering \includegraphics[width=8.5cm]{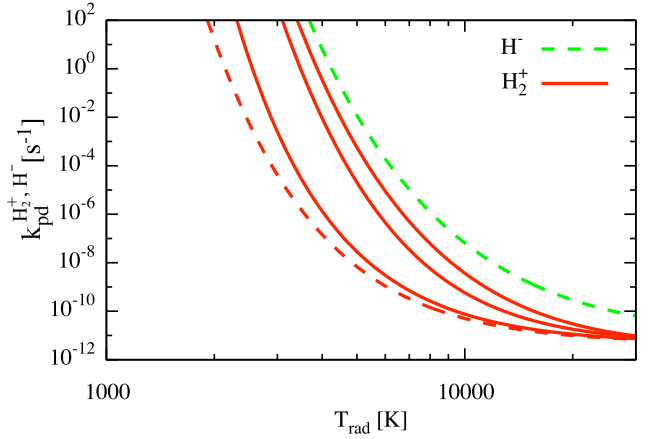}
    \caption{Photodissociation rate of $\mr{H_2^+}$ from the ground vibrational
    state (dashed red curve) and assuming LTE level populations with $T_{\rm rad}=1000, 3000$ and $8000$~K
    (solid red curves, bottom to top).  The dashed green curve shows the
    $\mr{H^-}$ photodetachment rate.  
    The radiation spectrum is a diluted black-body
    normalized as $J_{21}=1$.}
     \label{fig:kpdh2p}
\end{figure}
%overview of chemistry

In this section, we review our vibrationally-resolved chemical
network based on SOI14 and C11.  We consider 
29 chemical species ($\mr{H}$, $\mr{H}^+$, $\mr{H}^-$, $\mr{H_2}$,
$\mr{H_2^+}(v)$ with $v=0,\,1\,\ldots, 18$, $\mr{He}$, $\mr{He}^+$,
$\mr{He}^{2+}$, $\mr{HeH}^+$, $\mr{H_3}^+$ and $\mr{e}$), 
%resolving 19 different vibrational levels of $\mr{H_2^+}$, 
and we compute the evolution of
the fractional abundance of species $i$,
%with respect to the hydrogen atoms, 
$y(i)\equiv n(i)/n_\mr{H}$, where $n(i)$ is the number density of
species $i$ and $n_\mr{H}$ the number density of hydrogen nuclei.  We do
not consider deuterium chemistry because it should 
%not affect our conclusion.  Deuterium chemistry 
only affect  the evolution of the gas with
$T_\mr{gas}$ below a few hundred kelvin by HD cooling, a temperature regime that is never attained in 
our calculations 
\citep[see, e.g.,][]{Nagakura:2005aa,McGreer:2008aa,Wolcott-Green:2011ab,Nakauchi:2014aa}.

%reaction related to H2+
The processes involving $\mr{H_2^+}$ are the following.  The
vibrationally resolved reactions in our code are  $\mr{H_2^+}$ formation
by radiative association (reaction~\ref{eq:16}), $\mr{H_2^+}$
photodissociation (reaction~\ref{eq:3}), $\mr{H_2^+}$ dissociation by
charge transfer (reaction~\ref{eq:18}), vibrational
excitation/de-excitation by collision with $\mr{H}$ and vibrational
de-excitation by spontaneous emission.  We do not resolve
rotational levels of $\mr{H_2^+}$, because the dependence of the
efficiency of $\mr{H_2^+}$ photodissociation on rotational levels is
weaker than on vibrational levels \citep{Dunn:1968aa, Babb:2014aa}.
In addition, complete rovibrational state-resolved data are not available in
the literature.  Note that $\mr{H_2^+}$ formation via $\mr{HeH^+}$
hardly affects the evolution of primordial clouds (see, however, GP98 or C11 for its
consequences on the chemistry of the early Universe).

%implementation
The state-resolved data adopted in our chemical network are the following.
\cite{Babb:2014aa} provides rovibrationally resolved data for radiative
association and $\mr{H_2^+}$ photodissociation, which are summed up with
respect to rotational levels to obtain the vibrationally resolved rate
coefficients.  To obtain vibrationally resolved $\mr{H_2^+}$
photodissociation rate we assume rotational level population as
explained in Appendix~\ref{sec:rotational_assumption}.  
Fig.~\ref{fig:kpdh2p} shows the $\mr{H_2^+}$ photodissociation rates
$k_\mr{pd}^\mr{H_2^+}$ for a black-body radiation field and
$J_{21}=1$, obtained assuming (rovibrational) LTE population
with $T_{\rm rad}=1000$, 3000 and 8000$\,\mr{K}$, and all
$\mr{H_2^+}$ in the ground state.  Fig.~\ref{fig:kpdh2p} clearly shows that
$\mr{H_2^+}$ becomes more easily photodissociated as the $\mr{H_2^+}$
vibrational state becomes more excited.
Krsti\'c's database \citep{Krstic:2002aa,Krstic:2005aa,Krstic:1999aa,Krstic:2002aa,Krstic:2003aa}
provides vibrationally resolved data for charge transfer and 
vibrational excitation/de-excitation by collision with $\mr{H}$.
The rovibrationally resolved 
de-excitation rates of $\mr{H_2^+}$ by spontaneous emission are taken from
\cite{Posen:1983aa}. They are summed up with respect to rotational levels to obtain the
vibrationally resolved rates.
To see the dependence of the evolution of the clouds on the assumed $\mr{H_2^+}$ level
population, in addition to the standard state-resolved runs (hereafter {\em non-LTE model}),
we also perform runs using the (rovibrational) LTE $\mr{H_2^+}$ photodissociation rate (hereafter {\em LTE
model}) and runs using the $\mr{H_2^+}$ photodissociation rate for
$\mr{H_2^+}$ in the (rovibrational) ground state (hereafter {\em ground-state
model}). 

%H- and H2 photodissociation
We review here our treatment of $\mr{H^-}$ photodetachment and
indirect $\mr{H_2}$ photodissociation (the so-called Solomon process). 
Fig.~\ref{fig:kpdh2p} shows the $\mr{H^-}$
photodetachment rate coefficient $k_\mr{pd}^\mr{H^-}$ for a black-body
radiation field normalized to $J_{21}=1$ as a function of 
$T_\mr{rad}$. Here the cross section by \cite{John:1988aa} is used.  
As seen in Fig.~\ref{fig:kpdh2p}, the $\mr{H^-}$ photodetachment is generally 
more effective than $\mr{H_2^+}$ photodissociation although the latter 
depends on the $\mr{H_2^+}$ level population.
For the $\mr{H_2}$ photodissociation rate coefficient, 
we adopt the formula by \cite{Wolcott-Green:2011aa},
\begin{align}
k_\mr{pd}^\mr{H_2}= 1.4\times 10^{-12} f_\mr{sh}\, J_{21}\, \mr{s^{-1}}\,,
\label{eq:29}
\end{align}
where $f_\mr{sh}$ is the self-shielding factor \citep[for the explicit form,
see eq.~10 of][]{Wolcott-Green:2011aa}.  
For the radiation spectra considered in
Fig.~\ref{fig:jcrit},
Eq.~\eqref{eq:29} gives
$k_\mr{pd}^\mr{H_2}=1.4\times10^{-12}\,\mr{s^{-1}}$ without
self-shielding ($f_\mr{sh}=1$).  The fact that $k_\mr{pd}^\mr{H_2^+}$ and
$k_\mr{pd}^\mr{H^-}$ are decreasing functions of $T_\mr{rad}$ can be
understood as the  photodissociation of $\mr{H_2^+}$ and the 
photodetachment of $\mr{H^-}$ are relatively more effective for softer spectra,
compared to the  photodissociation of $\mr{H_2}$.  Note that the exact form of
$f_\mr{sh}$ is not settled in the literature \citep[see, e.g.,][]{Draine:1996aa,Wolcott-Green:2011aa,
Richings:2014aa,Hartwig:2015aa},
which may introduce an uncertainty in the value of $J_\mr{21}^\mr{cr}$ by a factor of a
few, as pointed out in SOI14 and \cite{Glover:2015ab}.  Similarly, evaluating
$k_\mr{pd}^\mr{H_2}$ with the intensity at a single frequency
$h\nu=12.4\,\mr{eV}$, may introduce an uncertainty in the value of $J_\mr{21}^\mr{cr}$
by a factor of a few if the spectrum changes substantially within the
LW bands, as mentioned in SOI14.
However, since the above uncertainties should not affect the conclusion of this
paper, we leave these issues for future studies.

%updated reaction, uncertainty
We updated the chemical network of SOI14, following
\cite{Glover:2015aa,Glover:2015ab}.  We include dissociative tunneling
of $\mr{H_2}$ to an unbound state \citep{Martin:1996aa} and $\mr{H}$
collisional ionization with $\mr{H}$
\citep{Lenzuni:1991aa,Gealy:1987aa} and $\mr{He}$
\citep{Lenzuni:1991aa,van-Zyl:1981aa}.  We replace the direct
$\mr{H_2}$ collisional dissociation rate with the fit given in \cite{Martin:1996aa}.

\subsection{Initial Conditions}
%initial condition
Following \cite{Omukai:2008aa} and SOI14, we take as initial
conditions of the calculation the following values corresponding to the
turnaround redshift $z=16$: $T_\mr{gas}=21\,\mr{K},\, n_\mr{H}= 4.5\times
10^{-3}\,\mr{cm^{-3}},\,y(\mr{e})=3.7\times10^{-4},\,
y(\mr{He})=8.3\times10^{-2},\,y(\mr{H_2})=2\times10^{-6}$ and $y(i)=0$
for the other species. Note that the initial chemical composition hardly affects the thermal evolution
of the clouds:  at the beginning of the evolution well within the
initial adiabatic contraction phase,
most of pre-existing $\mr{H_2}$ is photodissociated under a strong LW radiation field with $J_{21}$ around $J_{21}^\mr{cr}$;
$y(\mr{e})$ settles to the value determined by the condition that
the recombination time scale is about the same as the dynamical (free-fall) one;
and the abundances of the other species reach the values in chemical equilibrium \citep{Omukai:2001aa}.

%%%%%%%%%%%%%%%%%%%%%%%%%%%%%%%%%%%%%%%%%%%%%%%%%%%%%
%%%%%%%%%%%%%%%%%%%%%%%%%%%%%%%%%%%%%%%%%%%%%%%%%%%%%
\section{Results}
\label{sec:result}
%%%%%%%%%%%%%%%%%%%%%%%%%%%%%%%%%%%%%%%%%%%%%%%%%%%%%
%%%%%%%%%%%%%%%%%%%%%%%%%%%%%%%%%%%%%%%%%%%%%%%%%%%%%
In the following, we show how the temperature evolution 
of the cloud is modified by the assumed ${\rm H_2^+}$ vibrational 
level population. Then, we calculate the critical intensity of the radiation field $J_{21}^\mr{cr}$ as
a function of the black-body temperature 
and assess the importance of $\mr{H}_2$ formation via
the $\mr{H}_2^+$ channel on the evolution of the primordial clouds.

%%%%%%%%%%%%%%%%%%%%%%%%%%%%%%%%%%%%%%%%%%%%%%%%%%%%%
%%%%%%%%%%%%%%%%%%%%%%%%%%%%%%%%%%%%%%%%%%%%%%%%%%%%%
\subsection{Thermal evolution of the cloud}
\label{sec:evolution}
%%%%%%%%%%%%%%%%%%%%%%%%%%%%%%%%%%%%%%%%%%%%%%%%%%%%%
%%%%%%%%%%%%%%%%%%%%%%%%%%%%%%%%%%%%%%%%%%%%%%%%%%%%%
\begin{figure}
    \centering \includegraphics[width=8.5cm]{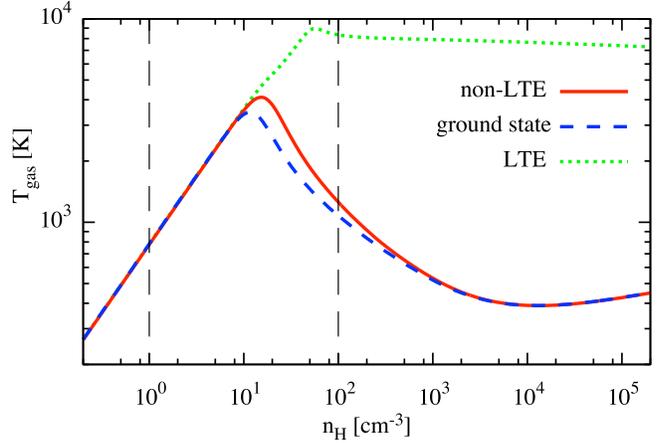} 
\caption{
Temperature evolution in a collapsing primordial gas cloud irradiated by a diluted black-body
radiation with $T_\mr{rad}=6000\,\mr{K}\mr{\ and \ } J_{21}=0.05$ 
as a function of the cloud density: non-LTE model (solid red), ground-state model (dashed blue) and LTE (dotted green).
The vertical lines (long-dashed black) demarcate boundaries between the low, intermediate and high-density regimes.}
\label{fig:nt} 
\end{figure}

\begin{figure}
    \centering \includegraphics[width=8.5cm]{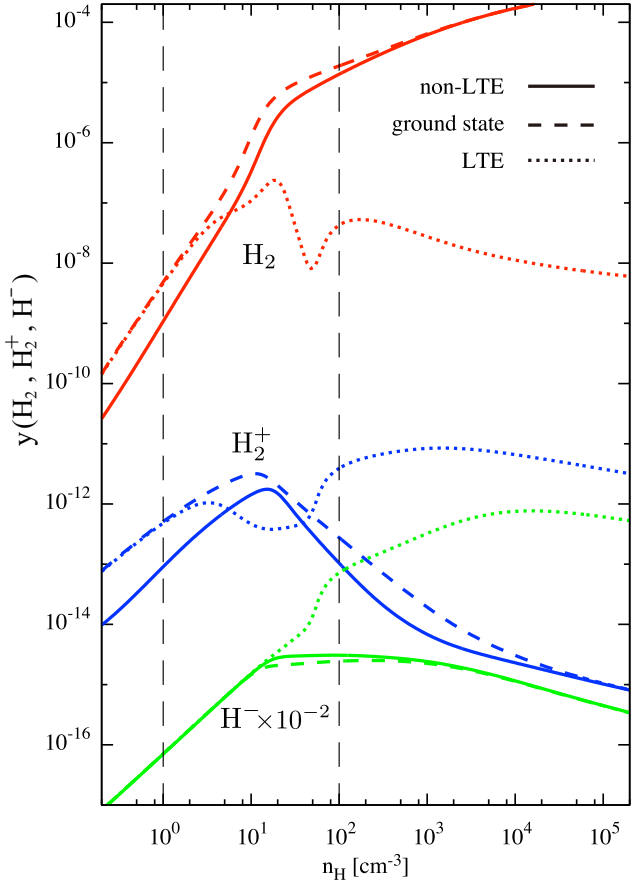}
    \caption{Abundances of $\mr{H_2}$ (red), $\mr{H_2^+}$ (blue) and $\mr{H^-}$
    ($\times 10^{-2}$; green) for the same cases as Fig.~\ref{fig:nt}: non-LTE model (solid), 
    ground-state model (dashed), and LTE (dotted) models.}  
    \label{fig:nfrac}
\end{figure}

\begin{figure}
\centering \includegraphics[width=8.5cm]{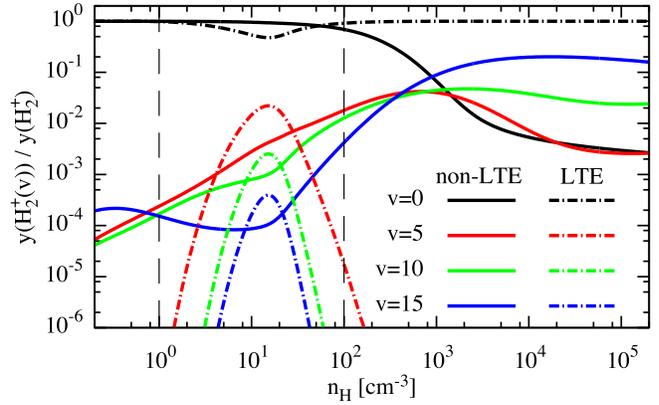}
\caption{Populations of $\mr{H_2^+}$ in the $v=0,5,10,$ and $15$
vibrational levels (black, red, green and blue, respectively) in
the non-LTE model (solid) for the same case as Fig.~\ref{fig:nt}.  The
LTE populations (dot dashed) with the temperature given by $T_\mr{gas}$
in the non-LTE model, i.e., the solid red curve in Fig.2, are also
plotted. }
\label{fig:poph2p}
\end{figure}

%one specific example
The run of the temperature, chemical abundance and $\mr{H_2^+}$ level
populations of a cloud irradiated by a black-body radiation field with
$T_\mr{rad}=6000\,\mr{K}$ and $J_{21}=0.05$ as a function of the
$\mr{H}$ number density are shown Figs.~\ref{fig:nt}, \ref{fig:nfrac}
and \ref{fig:poph2p}, respectively.  In Figs.~\ref{fig:nt} and
\ref{fig:nfrac} we also plot for comparison purposes the results for the
ground-state and LTE models, along with the fiducial results for the
non-LTE model.  In Fig.~\ref{fig:poph2p}, instead, we also plot the LTE
populations with the temperature given by $T_\mr{gas}$ in the non-LTE
model (i.e., the solid red curve in Fig.2) along with the results for
the non-LTE model, because the difference in $T_\mr{gas}$ in the non-LTE
and LTE models makes it difficult to extract the effect of non-LTE
chemistry by directly comparing the populations in these models.  The
results for the LTE model agree well with those in SOI14,  the
differences being due to the updated chemical network.  Below, we describe
the time (or, equivalently, density) evolution of the cloud with
particular attention to the $\mr{H_2^+}$ vibrational level population.

%begining
\subsubsection{Low-density Regime: $n_\mr{H}\lesssim 1\, \mr{cm^{-3}}$}
In this density range, the temperature is still very low, $T_\mr{gas}\lesssim
10^3\,\mr{K}$, although it rises with density by adiabatic compression, 
as seen in Fig.~\ref{fig:nt}.
Thus, almost all $\mr{H_2^+}$ is in the ground vibrational state
if LTE is assumed (Fig.~\ref{fig:poph2p}).
However, in the non-LTE model a finite amount of $\mr{H_2^+}$ is 
in higher vibrational levels because the $\mr{H_2^+}$ molecules 
tend to be formed by radiative association in excited states 
(reaction~\ref{eq:18}; \cite{Ramaker:1976aa,Babb:2014aa}).  
This effect is more pronounced at low temperatures where 
the $v=15$ level is more populated than the $v=5$ and $10$ levels.
Note that not all the $\mr{H_2^+}$ settles to the ground state 
even though the collisional excitation/de-excitation rate is much smaller 
than the radiative de-excitation rate in this low-density regime. 
This is because not only radiative de-excitation, but also 
radiative association and $\mr{H_2^+}$ photodissociation 
contribute to the $\mr{H_2^+}$ level population.
In the non-LTE model, 
since a fraction of $\mr{H_2^+}$ is in high vibrational levels, 
the larger $\mr{H_2^+}$ photodissociation reaction rate results in a
smaller $\mr{H_2^+}$ abundance than in the LTE or
ground-state models (Fig.~\ref{fig:nfrac}).  
As a result, the amount of $\mr{H_2}$ formed is smaller 
in the non-LTE case.
However, in this density range, the $\mr{H_2}$ abundance
is very low, $f_{H_2}\sim 10^{-9}$, in all models and 
does not affect the thermal evolution.

\subsubsection{Intermediate-density Regime: $n_\mr{H}\sim 1-10^2\, \mr{cm^{-3}}$}
\label{sec:mid_density}
%mid density
In this density regime, the evolutionary paths bifurcate into those
of atomic and $\mr{H_2}$ cooling depending on the assumed level population
(Fig.~\ref{fig:nt}).
As seen in Fig.~\ref{fig:poph2p}, 
at high temperature ($\sim$ 4000\,K), in the LTE approximation the
excited levels are somewhat populated 
around $\sim10\,\mr{cm^{-3}}$.
However, in the non-LTE treatment
only a smaller amount of $\mr{H_2^+}$ resides in excited states,
because of the higher photodissociation rate from those levels. 
In LTE the fractional abundance of excited levels, and thus
the $\mr{H_2^+}$ photodissociation rate are overestimated.
This results in smaller amounts of both $\mr{H_2^+}$ and 
$\mr{H_2}$, compared to the non-LTE model (Fig.~\ref{fig:nfrac}).  
Due to the lower H$_2$ fraction of the LTE case,
the cloud collapses along the atomic cooling track, 
whereas in non-LTE 
it follows the molecular hydrogen cooling path (Fig.~\ref{fig:nt}).  
Finally, in the ground-state model,
the smaller $\mr{H_2^+}$ photodissociation rate 
and the consequent larger amount of $\mr{H_2}$ (Fig.~\ref{fig:nfrac}) 
cause an earlier onset of
$\mr{H_2}$ cooling relative to the non-LTE case (Fig.~\ref{fig:nt}).

\subsubsection{High-density regime: $n_\mr{H}\gtrsim 10^2\, \mr{cm^{-3}}$}
%high density
The different evolutionary paths found in the previous regime are not modified at higher densities 
by the details of the various chemical processes.  Considering the large temperature 
difference between LTE and the other models, it is not possible to distinguish between the effects
of the level population and those related to the chemical abundances shown in Fig.~\ref{fig:nfrac}.
In this regime, unlike the LTE population, the excited levels are more populated than the ground state,
as seen in Fig.~\ref{fig:poph2p}.
The reason is as follows:
due to the high density, 
the $\mr{H_2^+}$ photodissociation is negligible compared to collisional
processes, such as radiative association, charge transfer and
collisional excitation/de-excitation.
Since not only collisional excitation/de-excitation, but also
radiative association and charge transfer 
continue to play important roles, 
the level population does not converge to the LTE value. This effect can be considered as a pumping
mechanism that depletes the ground level and allows for the occupation of higher excited states.

%%%%%%%%%%%%%%%%%%%%%%%%%%%%%%%%%%%%%%%%%%%%%%%%%%%%%
%%%%%%%%%%%%%%%%%%%%%%%%%%%%%%%%%%%%%%%%%%%%%%%%%%%%%
\subsection{The critical LW intensity $J_{21}^\mr{cr}$}
\label{sec:jcrit}
%%%%%%%%%%%%%%%%%%%%%%%%%%%%%%%%%%%%%%%%%%%%%%%%%%%%%
%%%%%%%%%%%%%%%%%%%%%%%%%%%%%%%%%%%%%%%%%%%%%%%%%%%%%
\begin{figure}
  \centering \includegraphics[width=8.5cm]{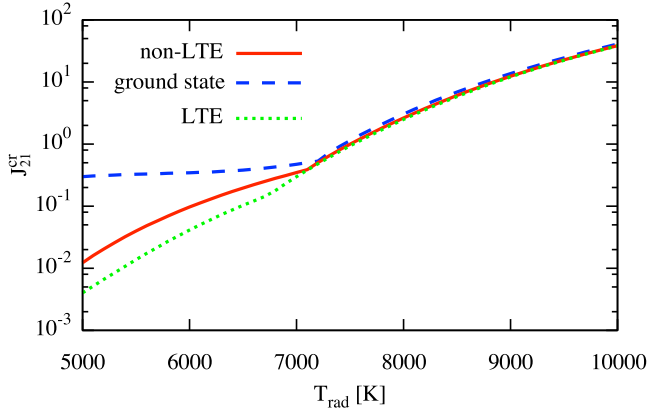} \caption{Critical intensity 
  for the direct collapse $J_{21}^\mr{cr}$ as a function of the black-body temperature
  $T_\mr{rad}$ of the irradiation radiation.  
  In addition to the value in the non-LTE model (solid red), 
  those in the ground-state (dashed blue) and
  LTE (dotted green) models are plotted.  }
  \label{fig:jcrit}
\end{figure}

%Jcr
Fig.~\ref{fig:jcrit} shows the critical LW intensity $J_{21}^\mr{cr}$ 
for the black-body spectra with temperature $T_\mr{rad}$.
We find $J_{21}^\mr{cr}$ by using the bisection method,
in which we examine whether the given $J_{21}$ is high enough for the cloud to
collapse along the atomic-cooling track by totally suppressing $\mr{H_2}$ cooling
(for more details, see, e.g., SOI14 and references therein). 
In addition to our fiducial result for the non-LTE model,
those for the ground-state and LTE models
are also plotted in Fig.~\ref{fig:jcrit} for comparison.

Two features in Fig.~\ref{fig:jcrit} should be noted.
%H2+ channel
First, the slope of $J_{21}^\mr{cr}$ changes discontinuously around
$T_\mr{rad}\sim 7000\,\mr{K}$, accompanying the shift of 
dominant $\mr{H_2}$-formation reaction determining $J_{21}^\mr{cr}$ 
from the $\mr{H_2^+}$ channel ($T_\mr{rad}\lesssim 7000\,\mr{K}$) to 
the $\mr{H^-}$ channel ($T_\mr{rad}\gtrsim
7000\,\mr{K}$).  We will discuss the condition for the
$\mr{H_2^+}$ channel to be dominant in Sec.~\ref{sec:influence}.
%population dependence
Second, in the range $T_\mr{rad}\lesssim 7000\,\mr{K}$, 
$J_{21}^\mr{cr}$ depends on the assumed $\mr{H_2^+}$ vibrational populations. 
Therefore, it is necessary to take into account the non-LTE $\mr{H}_2^+$ level population 
to obtain $J_{21}^\mr{cr}$ correctly because the efficiency of the
$\mr{H_2^+}$ channel, the dominant $\mr{H_2}$ formation channel in this range, 
is sensitive to its vibrational level population.
Compared to the correct $J_{21}^\mr{cr}$ by the non-LTE modelling, 
those for the ground-state and LTE models are
smaller or larger, respectively.  
This is because 
in the ground-state model
(the LTE model), compared to the non-LTE model, 
the $\mr{H_2^+}$ photodissociation rate 
is lower (higher) around the
density where the bifurcation into the two evolutionary paths occurs,
and thus more (less) $J_{21}$ is needed to suppress the $\mr{H_2}$ cooling.

%hard
On the other hand, $J_{21}^\mr{cr}$ for
$T_\mr{rad}\gtrsim7000\,\mr{K}$ is insensitive to the $\mr{H_2^+}$ level population 
because the $\mr{H}^-$ channel rather than the $\mr{H_2^+}$ channel 
is important in this range to determine $J_{21}^\mr{cr}$.  
Recall that typical high-$z$ galaxies have hard spectra corresponding to a black-body 
with $T_\mr{rad}\gtrsim2\times10^4\,\mr{K}$ (SOI14), as mentioned in
Sec.~\ref{sec:1-zone}.  Thus, our adoption of 
the LTE $\mr{H_2^+}$ photodissociation rate in SOI14 to obtain 
$J_{21}^\mr{cr}$ for realistic spectra is justified.  We plot
$J_{21}^\mr{cr}$ only for $T_\mr{rad}< 10^4\,\mr{K}$ in
Fig.~\ref{fig:jcrit}, because $J_{21}^\mr{cr}$ does not depend on 
the vibrational populations for $T_\mr{rad}>10^4\,\mr{K}$. 
For $J_{21}^\mr{cr}$ in the range $T_\mr{rad}>10^4\,\mr{K}$, 
we refer readers to,
e.g., Fig.3 of SOI14.

%%%%%%%%%%%%%%%%%%%%%%%%%%%%%%%%%%%%%%%%%%%%%%%%%%%%%
%%%%%%%%%%%%%%%%%%%%%%%%%%%%%%%%%%%%%%%%%%%%%%%%%%%%%
\subsection{
Condition for the $\mr{H_2^+}$-channel dominance
}
\label{sec:influence}
%%%%%%%%%%%%%%%%%%%%%%%%%%%%%%%%%%%%%%%%%%%%%%%%%%%%%
%%%%%%%%%%%%%%%%%%%%%%%%%%%%%%%%%%%%%%%%%%%%%%%%%%%%%

\begin{figure}
    \centering \includegraphics[width=8.5cm]{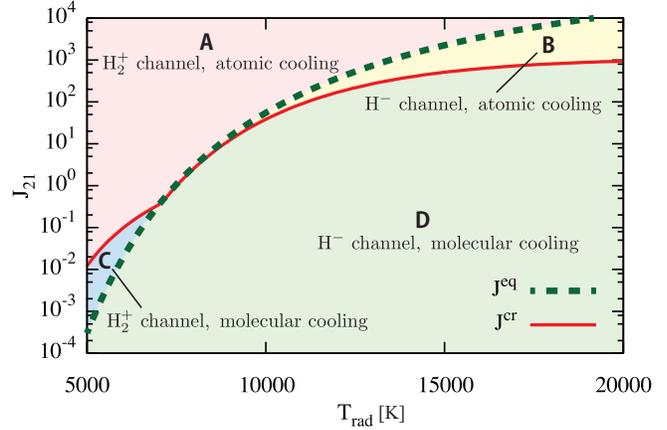} \caption{The
 threshold value $J_{21}^\mr{eq}$ (dashed green), given by
 Eq.~~\eqref{eq:1}, above which the $\mr{H_2^+}$ channel is more
 effective than the $\mr{H^-}$ channel at $n_\mr{H} =10^2\,
 \mr{cm^{-3}}$. The critical intensity $J_{21}^\mr{cr}$ for the direct collapse 
 shown in Fig.~\ref{fig:jcrit} (solid red) is
 overplotted. The explanation for the domains A-D is given in the
 text.  }  \label{fig:jeq}
\end{figure}

%this subsection
Here, we discuss the condition that 
the $\mr{H_2^+}$-channel dominates over the H$^{-}$ channel.
We introduce $J_{21}^\mr{eq}$, the threshold value of $J_{21}$ above which 
the $\mr{H_2^+}$ channel is more effective in H$_2$ formation than 
the $\mr{H^-}$ channel.

%H- channel

To begin with, we estimate the H$_2$ formation rate via the $\mr{H}^-$ channel.  
The H$^{-}$ fraction is determined by 
the balance between the radiative attachment (reaction~\ref{eq:7}),
associative detachment (reaction~\ref{eq:8}) and photodetachment
(reaction~\ref{eq:2}) as:
\begin{equation}
y(\mr{H^-})= \frac{k_\mr{ra}^\mr{H^-}\,y(\mr{e}) }{k_\mr{ad}^\mr{H^-}+
k_\mr{pd}^\mr{H^-}/n(\mr{H})}\,.
\label{eq:4}
\end{equation}
 The $\mr{H_2}$ formation rate via the $\mr{H}^-$
channel is given by
\begin{align}
\frac{{\rm d} y_\mr{H_2}^\mr{H^-}}{{\rm d}t}
&=k_\mr{ad}^\mr{H^-}\,y(\mr{H^-})\,n(\mr{H}) \nnmb
&\simeq \frac{ k_\mr{ra}^\mr{H^-}\,y(\mr{e})\,n(\mr{H})}
{1+k_\mr{pd}^\mr{H^-}/\,(n(\mr{H})\,k_\mr{ad}^\mr{H^-})}
\,,
\label{eq:26}      
\end{align}
where we have used Eq.~\eqref{eq:4} in the second line.
Similarly, we estimate the H$_2$ formation rate via the $\mr{H_2^+}$ channel.
The ${\rm H_2^+}$ fraction is determined by the balance between 
radiative association (reaction~\ref{eq:16}), charge
transfer (reaction~\ref{eq:18}) and photodissociation (reaction~\ref{eq:3}) as:
\begin{equation}
y(\mr{H_2^+}) = \frac{k_\mr{ra}^\mr{H_2^+}\,y(\mr{H^+}) }{k_\mr{ct}^\mr{H_2^+}+
k_\mr{pd}^\mr{H_2^+}/n(\mr{H})}\,. 
\label{eq:5}
\end{equation}
The $\mr{H_2}$ formation rate via the $\mr{H_2^+}$ channel, i.e., 
the charge transfer rate, is given by
\begin{align} 
\frac{{\rm d} y_\mr{H_2}^\mr{H_2^+}}{{\rm d}t}&=
k_\mr{ct}^\mr{H_2^+}\,y(\mr{H_2^+})\,n(\mr{H}) \nnmb
      &\simeq 
\frac{k_\mr{ra}^\mr{H_2^+}\,y(\mr{H^+})\,n(\mr{H})}
{1+k_\mr{pd}^\mr{H_2^+}/\,(n(\mr{H})\,k_\mr{ct}^\mr{H_2^+})}\,,
      \label{eq:27}
\end{align}
where we have used Eq.~\eqref{eq:5} in the second line.
By equating Eqs.~\eqref{eq:26} and \eqref{eq:27}, 
we obtain 
  \begin{align}
	J_{21}^\mr{eq}\simeq
\left(\frac{k_\mr{ra}^\mr{H^-}}{k_\mr{ra}^\mr{H_2^+}}-1\right)
\frac{k_\mr{ad}^\mr{H^-}\, n_\mr{H}}{\,k_\mr{pd}^\mr{H^-}|_{J_{21}=1}}\,,
       \label{eq:1}
      \end{align}
where we have used
$k_\mr{pd}^\mr{H^-}=k_\mr{pd}^\mr{H^-}|_{J_{21}=1}\times J_{21}$,
$n(\mr{H})\simeq n_\mr{H}$ and $y(\mr{e})\simeq y(\mr{H^+})$ and
neglected $k_\mr{pd}^\mr{H_2^+}$ for simplicity.  The effect of
$\mr{H_2^+}$ photodissociation would substantially change the
expression given by Eq.~\eqref{eq:1} if the LTE rate was used.  However,
the actual rate of $\mr{H_2^+}$ photodissociation just before the
atomic- and $\mr{H_2}$-cooling tracks bifurcate is lower than the LTE
one, as shown in Sec.~\ref{sec:mid_density}. Thus, 
the correction due to $\mr{H_2^+}$ photodissociation makes
$J_{21}^\mr{eq}$ given by Eq.~~\eqref{eq:1} larger at most by a factor of
a few and can be neglected in the order-of-magnitude estimate here.

%Jeq

To evaluate $J_{21}^\mr{eq}$, we consider the physical condition of the gas
just before the atomic- and $\mr{H_2}$-cooling tracks bifurcate (see,
e.g., \cite{Omukai:2001aa} or SOI14) and take $T_\mr{gas}=8000\,\mr{K}$
and $n_\mr{H}= 10^{2}\,\mr{cm^{-3}}$.  The density $n_\mr{H}$ at
bifurcation depends on the spectra and it is closer to $10^{2}\,\mr{cm^{-3}}$
for soft spectra with $T_\mr{rad}\lesssim 10^4\,\mr{K}$, which is of our
interest in this paper, although it is about $10^{3}\,\mr{cm^{-3}}$ for
$T_\mr{rad}\sim 10^5\,\mr{K}$.  The reaction rate coefficients at
$8000\,\mr{K}$ are given by $k_\mr{ra}^\mr{H^-}\approx
6\times10^{-15}\,\mr{cm^{3}\, s^{-1}}$, $k_\mr{ad}^\mr{H^-}\approx
9\times10^{-10}\,\mr{cm^{3}\, s^{-1}}$ and $k_\mr{ra}^\mr{H_2^+}\simeq
1.5\times 10^{-16}\,\mr{cm^{3}\, s^{-1}}$ 
(GP98) and the $\mr{H^-}$ photodetachment rate is given by
Fig.~\ref{fig:kpdh2p}.
%Figure
By substituting the above reaction rate coefficients and photodetachment rate
into Eq.~~\eqref{eq:1}, we
obtain $J_{21}^\mr{eq}$, as shown in Fig.~\ref{fig:jeq}.  Also shown is
$J_{21}^\mr{cr}$ obtained in Sec.~\ref{sec:jcrit} for comparison.
Depending on the value of $J_{21}$ relative to $J_{21}^\mr{cr}$ and
$J_{21}^\mr{eq}$, the way in which a cloud evolves falls within one of the
following four cases (Fig.~\ref{fig:jeq}):
\begin{itemize}
 \item[A:] $J_{21}>J_{21}^\mr{cr}$ and $J_{21}>J_{21}^\mr{eq}$.
The cloud collapses along the atomic-cooling track,
since $\mr{H_2}$, formed mainly via the $\mr{H_2^+}$ channel, is not enough to cool the gas.
 \item[B:] $J_{21}>J_{21}^\mr{cr}$ and $J_{21}<J_{21}^\mr{eq}$.
The cloud collapses along the atomic-cooling track,
since $\mr{H_2}$, formed mainly via the $\mr{H^-}$ channel, is not enough to cool the gas.
% for the same reasons described for Case A.
 \item[C:] $J_{21}<J_{21}^\mr{cr}$ and $J_{21}>J_{21}^\mr{eq}$.
The cloud collapses along the H$_2$-cooling track 
due to $\mr{H_2}$ formation mainly via the $\mr{H_2^+}$ channel.
 \item[D:]  $J_{21}<J_{21}^\mr{cr}$ and $J_{21}<J_{21}^\mr{eq}$.
The cloud collapses along the H$_2$-cooling track 
due to $\mr{H_2}$ formation mainly via the $\mr{H^-}$ channel.
\end{itemize}
In the soft-spectrum regime of $T_\mr{rad}\lesssim 7000\,\mr{K}$, where
$J_{21}^\mr{eq} \lesssim J_{21}^\mr{cr}$, the cases A, C, and D can be
realized depending on $J_{21}$.  In the case C, in particular, while the
$\mr{H^-}$ channel is blocked by radiation, sufficient $\mr{H_2}$ for
cooling can still form via the $\mr{H_2^+}$ channel.  
In this case, the proper account of the $\mr{H}_2^+$ level population is 
indispensable since the $\mr{H_2}$ formation rate via the $\mr{H}_2^+$ channel 
is sensitive to its level population, as shown by the example in
Sec.~\ref{sec:evolution}.
On the other hand,
in the hard-spectrum regime of $T_\mr{rad}\gtrsim 7000\,\mr{K}$, where
$J_{21}^\mr{cr} \lesssim J_{21}^\mr{eq}$, the cases A, B and D are
allowed: there is no range of $J_{21}$ satisfying
$J_{21}^\mr{eq}<J_{21}<J_{21}^\mr{cr}$.  Namely, radiation fields strong
enough to block the $\mr{H^-}$ channel always exceed the critical
intensity to totally suppress $\mr{H_2}$ formation by $\mr{H_2}$ 
photodissociation.  This means that the H$_2$-cooling track cannot be realized 
by the $\mr{H}_2^+$ channel for $T_\mr{rad}\gtrsim 7000\,\mr{K}$.

%%%%%%%%%%%%%%%%%%%%%%%%%%%%%%%%%%%%%%%%%%%%%%%%%%%%%
%%%%%%%%%%%%%%%%%%%%%%%%%%%%%%%%%%%%%%%%%%%%%%%%%%%%%
\section{ Discussion and Conclusions}
\label{sec:discussion}
%%%%%%%%%%%%%%%%%%%%%%%%%%%%%%%%%%%%%%%%%%%%%%%%%%%%%
%%%%%%%%%%%%%%%%%%%%%%%%%%%%%%%%%%%%%%%%%%%%%%%%%%%%%
%what is done in this paper
We have computed the thermal and chemical evolution of
primordial gas clouds resolving the vibrational levels of $\mr{H_2^+}$,
which enables us to properly implement $\mr{H_2}$ formation via the
$\mr{H_2^+}$ channel.  The efficiency of $\mr{H_2^+}$ photodissociation,
which suppresses the $\mr{H_2^+}$ channel by destroying the intermediate
product $\mr{H_2^+}$, is sensitive to the level population of the molecular ion.  We
have found that 
H$_2$ formation via the $\mr{H_2^+}$ channel becomes 
more effective in the non-LTE model than in LTE 
because more $\mr{H_2^+}$ is in the ground state and thus its
photodissociation rate is smaller.
%result (effect of the $\mr{H_2^+}$ channel)

As to the effects of the background radiation, we have found that 
$\mr{H_2}$ formation via the $\mr{H_2^+}$ channel becomes 
important in clouds irradiated by strong radiation fields with 
soft spectra characterised by $T_\mr{rad}\lesssim7000\,\mr{K}$.
In this case, the cloud thermal evolution is strongly affected by 
the $\mr{H_2^+}$ level population, indicating the importance of non-LTE 
treatment of the $\mr{H}_2^+$ vibrational levels.
On the other hand, under radiation fields with harder
spectra and $T_\mr{rad}\gtrsim7000\,\mr{K}$, 
the $\mr{H_2^+}$ channel always falls short of the $\mr{H^-}$ channel 
in producing H$_2$ molecules.

%result (Jcr)
We have derived
the critical radiation intensity $J_{21}^\mr{cr}$ 
for the formation of supermassive stars by direct collapse
and examined its dependence on the assumed $\mr{H_2^+}$ level population.  
For soft spectra with $T_\mr{rad}\lesssim 7000\,\mr{K}$,  
the critical intensity $J_{21}^\mr{cr}$ is under(over)-estimated if the 
level population is assumed to be in LTE 
(if all the $\mr{H_2^+}$ is assumed to be in the ground state). 
On the other hand, the value of $J_{21}^\mr{cr}$ is independent of
the assumed $\mr{H_2^+}$ level population for harder spectra with
$T_\mr{rad}\gtrsim 7000\,\mr{K}$.  Therefore, since typical 
high-redshift radiation sources, i.e., young and/or metal-poor galaxies, 
have rather hard spectra corresponding to black-bodies
with $T_\mr{rad}\gtrsim 2\times10^4\,\mr{K}$ (SOI14), 
the LTE approximation adopted in SOI14 in determining 
$J_{21}^\mr{cr}$ for realistic spectra is justified.

%possibility of soft spectra
We note that primordial gas clouds may be exposed to much softer radiation 
fields. For example, galaxies with extremely strong Ly-$\alpha$ emission
can be regarded as characterised by very soft spectra. Since the energy of
Ly-$\alpha$ photons ($h\nu=10.2\,\mr{eV}$) is below the lower limit
of the Lyman-Werner bands ($h\nu=11.2-13.6\,\mr{eV}$), such emission
would only destroy $\mr{H^-}$ leaving $\mr{H_2}$ unaffected.
Also, since the initial mass function in high-redshift galaxies is
poorly constrained, the infrared/optical cosmic background radiation 
originating from these sources can be soft with temperature 
$T_\mr{rad}\sim 6000\,\mr{K}$, corresponding to the typical
stellar mass $\sim 0.7\,M_\odot$, as considered in
\cite{Wolcott-Green:2012aa}.

In non-standard cosmology primordial density fluctuations are expected to be enhanced at
small scales, so that the first stellar objects can form at very early times 
with $z\gtrsim 100$, as considered in \cite{Hirano:2015ab}.
Under the strong and very soft CMB radiation at that time ($T_\mr{rad}\gtrsim 300\, \mr{K}$), 
H$_2$ formation via the ${\rm H_2^+}$ channel can be the dominant path with 
the $\mr{H^-}$ channel blocked by radiation processes  \citep[including non-thermal photons 
that play an additional role in enhancing H${^-}$ photodetachment;][]{CoppolaChluba2013}.

As a last example, let us consider the case where 
 direct collapse via the atomic-cooling track is 
realized. In such a case, the growing protostar results in a 
very extended structure with radius $R\simeq30\,\mr{AU}$ (``super-giant'' protostar) 
and low effective temperature 
$T_\mr{rad}\simeq 5000\,\mr{K}$ \citep{Hosokawa:2012aa,Hosokawa:2013aa}.  
In studying the physical conditions in the region around the super-giant protostar, 
the non-LTE $\mr{H_2^+}$ level population must be taken into account.

%future work
In this paper, the collapse of primordial gas clouds has been calculated 
by resolving the $\mr{H_2^+}$ vibrational levels, to our knowledge, for the first time.
However, neither $\mr{H_2^+}$ rotational levels 
nor $\mr{H_2}$ rovibrational levels have been explicitly included. 
Although the $\mr{H}_2^+$ rotationally-resolved kinetics is probably
not important, the $\mr{H}_2$ rovibrationally-resolved kinetics 
may have considerable effects.
The $\mr{H}_2$ rovibrational level population affects critical processes in the 
evolution of the primordial gas, 
such as photodissociation, 
collisional dissociation, and cooling by molecular hydrogen. 
We plan to include the full level-resolved kinetics of $\mr{H}_2$ and $\mr{H}_2^+$ in future studies.
\\

KS would like to thank D. Nakauchi, H. Yajima and K. Toma for fruitful
 discussions and valuable comments.  
KS would also like to thank 
Savino Longo and University of Bari 
for hospitality and support during his visit to University of Bari.
This work is supported in part by
 MEXT/JSPS KAKENHI Grant Number 15J03873 (KS) and 25287040 (KO). The authors 
 acknowledge the discussions within the international team \#272 lead by
C. M. Coppola ``EUROPA - Early Universe: Research on Plasma Astrochemistry'' at ISSI
(International Space Science Institute) in Bern.

%\bibliography{$HOME/physics/mybib}

\appendix
%%%%%%%%%%%%%%%%%%%%%%%%%%%%%%%%%%%%%%%%%%%%%%%%%%%%%
%%%%%%%%%%%%%%%%%%%%%%%%%%%%%%%%%%%%%%%%%%%%%%%%%%%%%
\section{$\mr{H_2^+}$ rotational level population }
\label{sec:rotational_assumption}
%%%%%%%%%%%%%%%%%%%%%%%%%%%%%%%%%%%%%%%%%%%%%%%%%%%%%
%%%%%%%%%%%%%%%%%%%%%%%%%%%%%%%%%%%%%%%%%%%%%%%%%%%%%
\begin{figure}
    \centering \includegraphics[width=8.5cm]{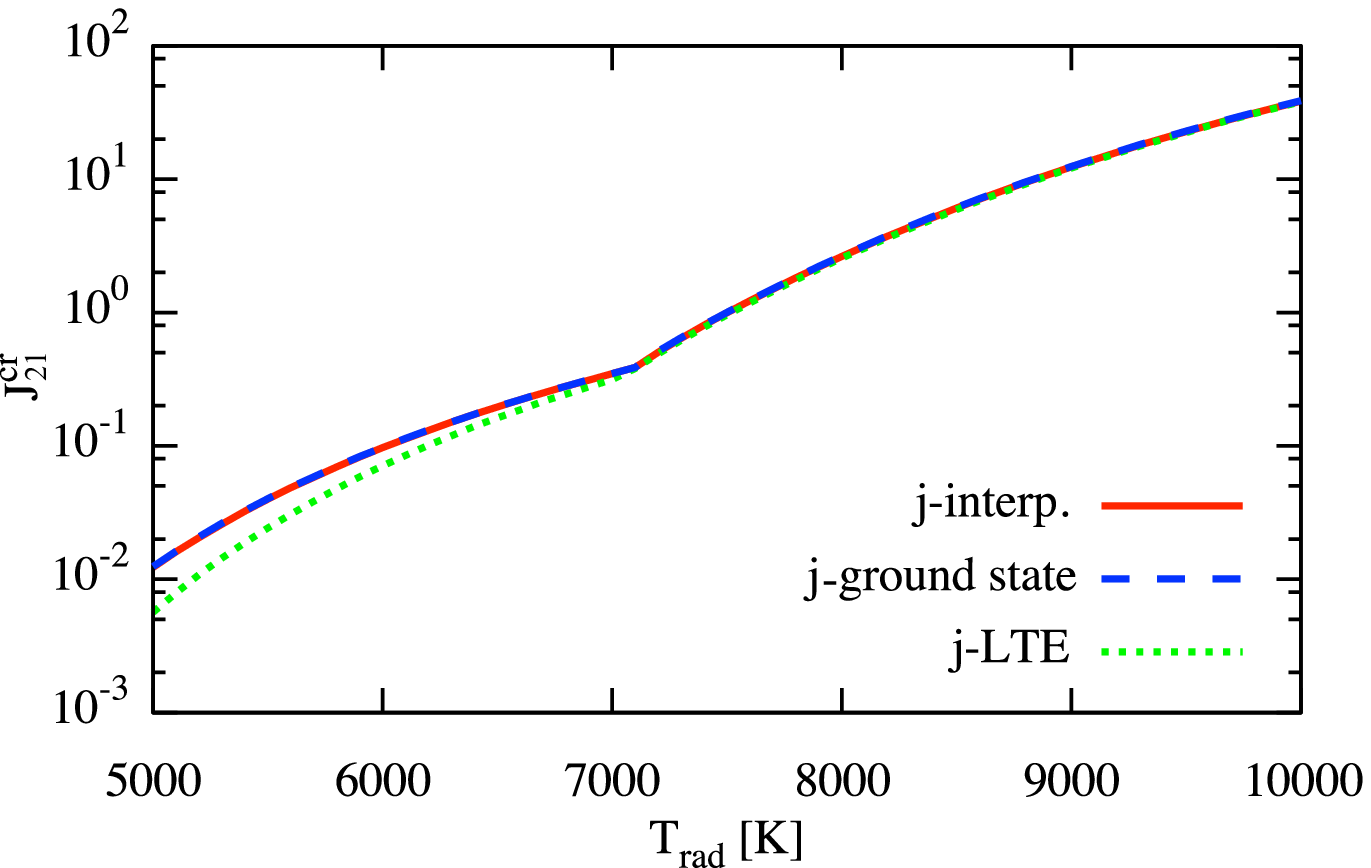}
     \caption{Same as Fig.~\ref{fig:jcrit}, but for three different assumptions on 
 the $\mr{H_2^+}$ rotational level population: ({\em i}\/) model using the interpolation formula 
%by \cite{Glover:2015aa},
by Glover (2015a),
({\em ii}\/) model where all the $\mr{H_2^+}$ in each vibrational level is 
 in its rotational ground state. 
 and ({\em iii}\/) model assuming the rotational LTE. 
 Note that the $\mr{H_2^+}$ vibrational levels are resolved in all the cases.}
\label{fig:jcrit_rotdep}
\end{figure}

Here, we describe our treatment of the $\mr{H_2^+}$ rotational level
populations.
Since comprehensive rotational-state-resolved data is not available in the
literature, we use the interpolation formula by \cite{Glover:2015aa}.
Specifically, we calculate the $\mr{H_2^+}$ photodissociation rate from
each vibrational level by the formula
$k=k_\mr{LTE}\,\left(k_\mr{n\to0}/k_\mr{LTE}\right)^\alpha$,
where $k_\mr{n\to0}$ and $k_\mr{LTE}$ are the rates for the ground
rotational state and the LTE rotational level population, respectively, 
$\alpha=(1+n_\mr{H}/n_\mr{cr})^{-1}$
and $n_\mr{cr}$ is the critical density for the LTE populations,
which is $\sim10^3\,\mr{cm^{-3}}$ for $T_\mr{gas}=8000\,\mr{K}$ \citep{Glover:2015aa}.

To estimate the error in our modeling of the $\mr{H_2^+}$ rotational
level population, we compare the critical intensity $J_{21}^\mr{cr}$ 
obtained with three different modeling of rotational level population: 
(i) the fiducial model based on \cite{Glover:2015aa} ($j$-interp. model); 
(ii) the model assuming that all the $\mr{H_2^+}$ in each vibrational level 
resides in the ground rotational state ($j$-ground
state model); and (iii) the model assuming the LTE rotational level population in
each vibrational level ($j$-LTE model).  Fig.~\ref{fig:jcrit_rotdep}
shows that the values of $J_{21}^\mr{cr}$ in the $j$-interp. and $j$-ground
state models are almost identical while in the $j$-LTE model it is slightly smaller.
This difference is, however, much smaller than that among 
the vibrationally non-LTE and LTE models shown in Fig.~\ref{fig:jcrit},
indicating that the cloud thermal evolution does not depend so much 
on rotational level population as the
vibrational level population.

\end{document}